\documentclass[conference]{IEEEtran}

\usepackage{amsmath, amsfonts} 
\usepackage{amsthm}
\usepackage[cmintegrals]{newtxmath}
\usepackage{graphicx}
\usepackage{multirow}
\usepackage{cases}

\ifCLASSOPTIONcompsoc
\usepackage[caption=false,font=normalsize,labelfont=sf,textfont=sf]{subfig}
\else
\usepackage[caption=false,font=footnotesize]{subfig}
\fi

\newcommand{\overbar}[1]{\mkern 1.5mu\overline{\mkern-1.5mu#1\mkern-1.5mu}\mkern 1.5mu}
\begin{document}

\title{Transversal GRAND for Network Coded Data}

\author{\IEEEauthorblockN{Ioannis~Chatzigeorgiou}
\IEEEauthorblockA{School of Computing and Communications\\
Lancaster University, United Kingdom\\
Email: i.chatzigeorgiou@lancaster.ac.uk}}

\maketitle
\begin{abstract}

This paper considers a transmitter, which uses random linear coding (RLC) to encode data packets. The generated coded packets are broadcast to one or more receivers. A receiver can recover the data packets if it gathers a sufficient number of coded packets. We assume that the receiver does not abandon its efforts to recover the data packets if RLC decoding has been unsuccessful; instead, it employs syndrome decoding in an effort to repair erroneously received coded packets before it attempts RLC decoding again. A key assumption of most decoding techniques, including syndrome decoding, is that errors are independently and identically distributed within the received coded packets. Motivated by the `guessing random additive noise decoding' (GRAND) framework, we develop transversal GRAND: an algorithm that exploits statistical dependence in the occurrence of errors, complements RLC decoding and achieves a gain over syndrome decoding, in terms of the probability that the receiver will recover the original data packets.
\vspace{2pt}
\end{abstract}



\section{Introduction}
\label{sec.Introduction}

Random linear coding (RLC)~\cite{Ho2006}, which encompasses fountain coding and network coding, is a forward error correction scheme that has the potential to increase throughput and improve resilience to network attacks. It was originally developed for layers higher than the physical layer, e.g., the network layer and the application layer. RLC encoding generates coded packets that are random linear combinations of input data packets. RLC decoding discards received coded packets that have been corrupted by errors and attempts to reconstruct the original data packets from correctly received coded packets.

When RLC decoding is unsuccessful, discarded corrupted packets can be utilized. Syndrome decoding~\cite{Mohammadi2016}, which focuses on binary transmission, aims to repair corrupted packets in polynomial time. Although syndrome decoding can improve the chances of the RLC decoder recovering the data packets, it does not account for correlations between errors. To address this deficiency, we look into the recently proposed `guessing random additive noise decoding' (GRAND) scheme.

GRAND is a code-agnostic decoding scheme that endeavors to identify and nullify the effect of noise on transmitted signals. The theoretical foundations and properties of GRAND were laid in \cite{Duffy2018} and \cite{Duffy2019}. A practical implementation of the GRAND framework, referred to as GRAND Markov order (GRAND-MO), was presented in \cite{An2020}. GRAND-MO is a hard detection decoder of binary codewords that have been impaired by errors correlated over time.

GRAND-MO focuses on physical-layer coding, according to which $N$-bit codewords are generated by $K$-bit input words, where $N\geq K$, and are transmitted over a noisy channel. As a result, correlated errors are distributed over the $N$ bits of each codeword. On the other hand, an RLC encoder organizes input bits into $K$ packets and generates $N$ coded packets, which are random linear combinations of the $K$ packets. In essence, a bit in position $b$ of a coded packet is obtained by the randomly weighted modulo-$2$ addition of the bits that occupy position $b$ in all of the $K$ input packets. Therefore, the bits in position $b$ of the $N$ coded packets create an $N$-bit codeword. Equivalently, if the coded packets are stacked up and form the rows of a matrix, the codewords will occupy the columns of this matrix. Correlated errors introduced by the channel will be present in each coded packet. This means that errors across adjacent codewords will be correlated but errors within each codeword will be uncorrelated. The objective of this paper is to build on the concept of GRAND and propose \textit{transversal GRAND}, an algorithm that exploits correlations between errors across multiple codewords -- and is thus suitable for assisting RLC decoding -- in contrast to GRAND-MO, which leverages correlations between errors within codewords.

The remainder of this paper has been organized as follows: Section~\ref{sec.RLC} briefly describes RLC decoding, which discards packets received in error. Section~\ref{sec.GRAND} explains how syndrome decoding can repair packets in error and assist RLC decoding. Transversal GRAND is then introduced as a means to exploit the statistical structure of errors and improve the chances of repairing packets. Performance comparisons between syndrome decoding and transversal GRAND are presented in Section~\ref{sec.Results} and key findings are summarized in Section~\ref{sec.Conclusions}.


\section{Random Linear Coding and Decoding}
\label{sec.RLC}

We consider a source node that intends to broadcast messages to one or more destination nodes. Before initiating transmission, the source node segments each message into $K$ packets. Each source packet has been modeled as a sequence of $B$ bits. The $K$ source packets of $B$ bits can be expressed as a matrix $\mathbf{U}\in\mathbb{F}^{K\times B}_{2}$, where $\mathbb{F}^{K\times B}_{2}$ denotes the set of all $K\times B$ matrices over $\mathbb{F}_{2}=\{0,1\}$. RLC~\cite{Ho2006} is used to encode the $K$ source packets into $N\geq K$ coded packets. The $N$ coded packets can also be expressed in matrix form as $\mathbf{X}\in\mathbb{F}^{N\times B}_{2}$. The relationship between matrices $\mathbf{X}$ and $\mathbf{U}$ is:
\begin{equation}
\label{eq.RLNC_matrix}
\mathbf{X} = \mathbf{G}\,\mathbf{U},
\end{equation}
where $\mathbf{G}$ is known as the \textit{generator matrix}. We employ \textit{systematic} RLC encoding~\cite{Jones15}, according to which the first $K$ of the $N$ transmitted packets are identical to the $K$ source packets, while the remaining $N-K$ coded packets are random linear combinations of the source packets. In systematic RLC, the generator matrix $\mathbf{G}$ can be expressed in standard form:
\begin{equation}
\label{eq.sys_generator_matrix}
\mathbf{G}=
\left[\!\begin{array}{c} \mathbf{I}_K\\ \mathbf{P}\end{array}\!\right],
\end{equation}
where $\mathbf{I}_K$ is the $K\times K$ identity matrix. Each element of the $(N-K)\times K$ matrix $\mathbf{P}$ is chosen uniformly and at random from $\mathbb{F}_2$. As explained in \cite{Mohammadi2016}, a seed could be used to initialize the pseudo-random number generator that outputs the elements of $\mathbf{P}$. Given that the value of the seed can be conveyed in the headers of the coded packets, we assume that the destination nodes have knowledge of $\mathbf{G}$.

Let $\mathbf{Y}$ be an erroneous copy of $\mathbf{X}$ that has been received by a destination node, defined as:
\begin{equation}
\label{eq.Y_at_destination}
	\mathbf{Y}=\mathbf{X}\oplus \mathbf{E},
\end{equation}
where $\oplus$ represents modulo-$2$ addition. The \textit{error matrix} $\mathbf{E}$ contains ones in positions where errors have occurred and zeros in the remaining positions. The destination node classifies the received coded packets into error-free and erroneous coded packets, e.g., using cyclic redundancy checks (CRCs). Let $\mathcal{R}$ be the set of the row indices of $\mathbf{Y}$ that correspond to correctly received coded packets, and $N_\mathrm{R}$ be the number of coded packets that contain no errors, i.e., $\vert \mathcal{R}\vert=N_\mathrm{R}\leq N$. The destination node constructs matrix $\mathbf{Y}_\mathcal{R}$ from the $N_\mathrm{R}$ correctly received coded packets and matrix $\mathbf{G}_\mathcal{R}$ from those $N_\mathrm{R}$ rows of $\mathbf{G}$ with indices in $\mathcal{R}$. On the other hand, the $N-N_\mathrm{R}$ erroneously received coded packets have indices in the set $\overbar{\mathcal{R}}=\{1,\ldots,N\}\backslash \mathcal{R}$ and make the rows of matrix $\mathbf{Y}_{\overbar{\mathcal{R}}}$. Using the indices in $\overbar{\mathcal{R}}$, matrices $\mathbf{X}_{\overbar{\mathcal{R}}}$, $\mathbf{G}_{\overbar{\mathcal{R}}}$ and $\mathbf{E}_{\overbar{\mathcal{R}}}$ are obtained. The relationships between these matrices are summarized below:
\begin{subnumcases}
{\mathbf{Y} = \mathbf{X}\oplus\mathbf{E}\Leftrightarrow}
      \mathbf{Y}_\mathcal{R}=\mathbf{X}_\mathcal{R}=\mathbf{G}_\mathcal{R}\mathbf{U},\label{eq.error_free}\\
      \mathbf{Y}_{\overbar{\mathcal{R}}}=\mathbf{X}_{\overbar{\mathcal{R}}}\oplus\mathbf{E}_{\overbar{\mathcal{R}}}\label{eq.erroneous}.
\end{subnumcases}

RLC decoding utilizes only the $N_\mathrm{R}$ correctly received coded packets represented by $\mathbf{Y}_\mathcal{R}$. The $N-N_\mathrm{R}$ erroneous coded packets contained in $\mathbf{Y}_{\overbar{\mathcal{R}}}$ are discarded. The source message, represented by matrix $\mathbf{U}$, can be obtained from $\mathbf{X}_\mathcal{R} = \mathbf{G}_\mathcal{R}\,\mathbf{U}$ if the rank of $\mathbf{G}_\mathcal{R}$ is $K$. In that case, the relationship $\mathbf{X}_\mathcal{R} = \mathbf{G}_\mathcal{R}\,\mathbf{U}$ can be seen as a system of $N_\mathrm{R}\geq K$ linear equations, which can be reduced to a system of $K$ linearly independent equations with $K$ unknowns, i.e., source packets. This $K\times K$ system can be solved, e.g., using Gaussian elimination, and return a unique solution for $\mathbf{U}$. Otherwise, if $\mathrm{rank}(\mathbf{G}_\mathcal{R})<K$, the linear system has fewer linearly independent equations than unknowns, hence a unique solution cannot be obtained.

If the decoder does not discard the erroneous coded packets but attempts to repair them, the probability of recovering the source message can potentially be increased. The following section describes syndrome decoding~\cite{Mohammadi2016}, which aims to repair coded packets that contain errors while operating under the assumption that transmission is over a binary symmetric channel (BSC). We then propose transversal GRAND, which takes into consideration the statistical properties of the channel and increases the chances of recovering the source message.


\section{Guessing Random Additive Noise}
\label{sec.GRAND}

Guessing random additive noise is the process of calculating an estimate of $\mathbf{E}_{\overbar{\mathcal{R}}}$, denoted by $\hat{\mathbf{E}}_{\overbar{\mathcal{R}}}$. An estimate of the transmitted coded packets can then be derived from \eqref{eq.erroneous}:
\begin{equation}
\label{eq.Estimated_X}
\hat{\mathbf{X}}_{\overbar{\mathcal{R}}}=\mathbf{Y}_{\overbar{\mathcal{R}}}\oplus\hat{\mathbf{E}}_{\overbar{\mathcal{R}}}.
\end{equation}
To determine which of the estimated rows of $\hat{\mathbf{X}}_{\overbar{\mathcal{R}}}$ correspond to successfully repaired coded packets, CRC verification is carried out. Let $\nu$ denote the number of rows in $\hat{\mathbf{X}}_{\overbar{\mathcal{R}}}$ that passed CRC verification. The indices of the $\nu$ repaired coded packets are removed from set $\overbar{\mathcal{R}}$ and are added to set $\mathcal{R}$, while the corresponding rows of $\hat{\mathbf{X}}_{\overbar{\mathcal{R}}}$ are moved to $\mathbf{Y}_\mathcal{R}$ in \eqref{eq.error_free}. The cardinalities of sets $\overbar{\mathcal{R}}$ and $\mathcal{R}$ change to $N-N_\mathrm{R}-\nu$ and $N_\mathrm{R}+\nu$, respectively, while the dimensions of $\hat{\mathbf{X}}_{\overbar{\mathcal{R}}}$ and $\mathbf{Y}_\mathcal{R}$ change to $(N-N_\mathrm{R}-\nu)\times B$ and $(N_\mathrm{R}+\nu)\times B$, respectively. The indices of the $\nu$ repaired packets are also used to identify the rows of the generator matrix $\mathbf{G}$ that should be appended to $\mathbf{G}_\mathcal{R}$ in \eqref{eq.error_free}. If the $\nu$ repaired packets increase the rank of the enlarged $(N_\mathrm{R}+\nu)\times K$ matrix $\mathbf{G}_\mathcal{R}$ to $K$, the process of guessing random additive noise will have been successful in assisting the RLC decoder to recover the $K$ source packets.

This section briefly describes syndrome decoding~\cite{Mohammadi2016}, which is an established method for computing $\hat{\mathbf{E}}_{\overbar{\mathcal{R}}}$. We explain the shortcomings of syndrome decoding when the errors in each coded packet are correlated, and we propose transversal GRAND, which builds on the concept of GRAND and exploits error correlations that can lead to error bursts.

\subsection{Syndrome Decoding}

Given that the $N\times K$ generator matrix $\mathbf{G}$ is known to all destination nodes, the $N\times (N-K)$ \textit{parity-check matrix} $\mathbf{H}$ can be derived as follows:
\begin{equation}
\label{eq.sys_parity_check_matrix}
\mathbf{H}=\left[\!\begin{array}{c} -\mathbf{P}\;\vert\;\mathbf{I}_{N-K}\end{array}\!\right]^\top,
\end{equation}
so that:
\begin{equation}
\label{eq.zero_product}
\mathbf{H}^\top\,\mathbf{G}=\mathbf{0}.
\end{equation}
When operations are performed in $\mathbb{F}_2$, negation has no effect on a matrix, i.e., $-\mathbf{P}=\mathbf{P}$. Note that $\mathbf{H}$ can be obtained from $\mathbf{G}$ even when RLC is non-systematic, as long as $\mathbf{G}$ has rank $K$; in that case, we can apply column-wise Gaussian elimination to transform $\mathbf{G}$ into standard form, as in \eqref{eq.sys_generator_matrix}, identify $\mathbf{P}$ and obtain $\mathbf{H}$ using \eqref{eq.sys_parity_check_matrix}.

Multiplication of $\mathbf{H}^\top$ by the received matrix $\mathbf{Y}$ produces the $(N-K)\times B$ \textit{syndrome matrix} $\mathbf{S}$, i.e., $\mathbf{S} = \mathbf{H}^\top \mathbf{Y}$. Using \eqref{eq.zero_product}, we find that the relationship between the syndrome matrix $\mathbf{S}$ and the error matrix $\mathbf{E}$ is:
\begin{equation}
\label{eq.syndrome_decoding_full}
\mathbf{S} = \mathbf{H}^\top \mathbf{Y}
 = \mathbf{H}^\top (\mathbf{X}\oplus\mathbf{E})
 = \mathbf{H}^\top (\mathbf{G}\mathbf{U}\oplus\mathbf{E})
 = \mathbf{H}^\top \mathbf{E}.
\end{equation}
As our focus is on the $N-N_\mathrm{R}$ erroneous coded packets, we use the set $\overbar{\mathcal{R}}$ to isolate the $N-N_\mathrm{R}$ of the $N$ rows of $\mathbf{H}$ and $\mathbf{E}$ in \eqref{eq.syndrome_decoding_full}, and construct $\mathbf{H}_{\overbar{\mathcal{R}}}$ and $\mathbf{E}_{\overbar{\mathcal{R}}}$, respectively. Consequently, expression \eqref{eq.syndrome_decoding_full} changes to:
\begin{equation}
\label{eq.syndrome_decoding_partial}
\mathbf{S} = \left(\mathbf{H}_{\overbar{\mathcal{R}}}\right)^{\!\top} \mathbf{E}_{\overbar{\mathcal{R}}}.
\end{equation}
If column $b$ of $\mathbf{S}$ and $\mathbf{E}_{\overbar{\mathcal{R}}}$ is denoted by $[\mathbf{S}]_{*,b}$ and $[\mathbf{E}_{\overbar{\mathcal{R}}}]_{*,b}$, respectively, expression \eqref{eq.syndrome_decoding_partial} can be re-written as $B$ independent systems of $N-K$ linear equations with $N-N_\mathrm{R}$ unknowns per equation:
\begin{equation}
\label{eq.syndrome_decoding_partial_per_col}
\left[\mathbf{S}\right]_{*,b} = \left(\mathbf{H}_{\overbar{\mathcal{R}}}\right)^{\!\top} \left[\mathbf{E}_{\overbar{\mathcal{R}}}\right]_{*,b}\quad\text{for}\quad b=1,\ldots,B.
\end{equation}

Mohammadi~\textit{et al.}~\cite{Mohammadi2016} observed that erroneously received coded packets usually contain large error-free segments, thus $\mathbf{E}_{\overbar{\mathcal{R}}}$ is a \textit{sparse} matrix, that is, most of the elements in $\mathbf{E}_{\overbar{\mathcal{R}}}$ are zero-valued. Based on this observation, the solution to \eqref{eq.syndrome_decoding_partial_per_col} can be formulated as:
\begin{IEEEeqnarray}{ll}
\label{eq.norm_minimization}
\left[\hat{\mathbf{E}}_{\overbar{\mathcal{R}}}\right]_{*,b}=\;  &\arg\min_{\mathbf{w}^{\top}}\;\lVert \mathbf{w} \rVert_0 \IEEEyesnumber\IEEEyessubnumber*\label{eq.obj_func}\\
&\text{subject to}\; \left(\mathbf{H}_{\overbar{\mathcal{R}}}\right)^{\!\top}\mathbf{w}^{\top}=\left[\mathbf{S}\right]_{*,b}
\label{eq.constraint}
\end{IEEEeqnarray}
where $\mathbf{w}\in\mathbb{F}^{N-N_\mathrm{R}}_2$ is a row vector that should satisfy constraint \eqref{eq.constraint} and have the minimum possible number of non-zero elements. The norm $\lVert \mathbf{w} \rVert_0$, which counts the non-zero elements in $\mathbf{w}$, is defined as $\lVert \mathbf{w} \rVert_0=\lvert w_1\rvert^0+\ldots+\lvert w_{N-N_\mathrm{R}}\rvert^0$ assuming that $0^0=0$ \cite{Donoho2001}. Syndrome decoding considers \eqref{eq.obj_func} and initiates an exhaustive search for a solution; the sparsity of the row vector $\mathbf{w}$ is gradually reduced, i.e., the number of ones in $\mathbf{w}$ increases, and the search concludes when the sparsest vector $\mathbf{w}$ that satisfies \eqref{eq.constraint} has been identified.

Although syndrome decoding has the potential to assist RLC decoding and increase the probability of recovering the source packets~\cite{Mohammadi2016}, it does not exploit the statistical properties of the channel. It selects columns for $\hat{\mathbf{E}}_{\overbar{\mathcal{R}}}$ that are as sparse as possible or, equivalently, have the lowest possible Hamming weight, but it does not account for the possibility that the columns of $\hat{\mathbf{E}}_{\overbar{\mathcal{R}}}$ might be correlated. Thus, syndrome decoding is suitable for memoryless channels, like the BSC, although channels often have memory in practice. The two-state Markov model~\cite{Gilbert1960} is a widely used channel model for the characterization of the correlation between consecutive errors.

\begin{figure}[t]
\centering
\includegraphics[width=0.52\columnwidth]{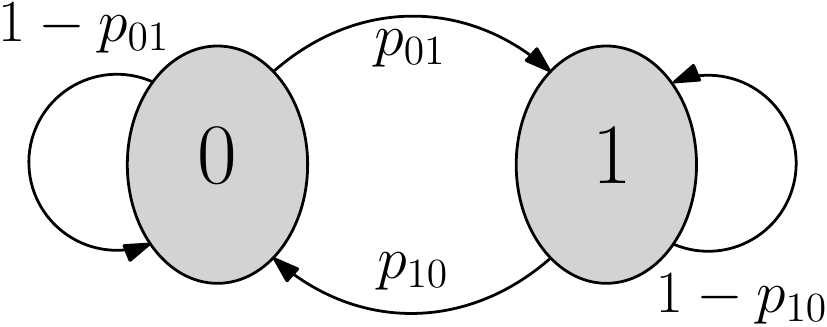}
\caption{The two-state Markov model, where $0$ and $1$ represent the `good' state and the `bad' state, respectively.}
\label{fig:channel_model}
\end{figure}

\subsection{Two-state Markov Model for Error Bursts}
\label{subsec:Gilbert}

As in \cite{An2020}, we use a two-state Markov chain to model error bursts. According to this model, a bit is received in error if a transition to the `bad' state of the chain, denoted by $1$, occurs. Otherwise, a transition to the `good' state, denoted by $0$, implies that the bit has been received correctly \cite{Gilbert1960}. Transitions between the two states occur with probabilities $p_{01}$, $1-p_{01}$, $p_{10}$ and $1-p_{10}$, as shown in Fig.~\ref{fig:channel_model}. The steady-state probability of being in state $1$ represents the bit error probability and is given by $\varepsilon=p_{01}/(p_{01}+p_{10})$ \cite{Rancurel2001,Cloud2015}. The expected number of consecutive errors provides the average length of an error burst, and is equal to $\Lambda=1/p_{10}$ \cite{Cloud2015}.

\subsection{Transversal GRAND}

In syndrome decoding, column vectors that have the same number of non-zero elements, i.e., the same Hamming weight, are treated as being equally likely and can thus be queried in any order by \eqref{eq.norm_minimization}. Furthermore, all column vectors of a given Hamming weight $w$ are queried before column vectors of weight $w+1$. The proposed method, which we call transversal GRAND, queries column vectors in descending order of likelihood, as dictated by the transition probabilities of the Markov model and not by the Hamming weights of the vectors.

\begin{figure*}[t]
\centering
\includegraphics[width=1.9\columnwidth]{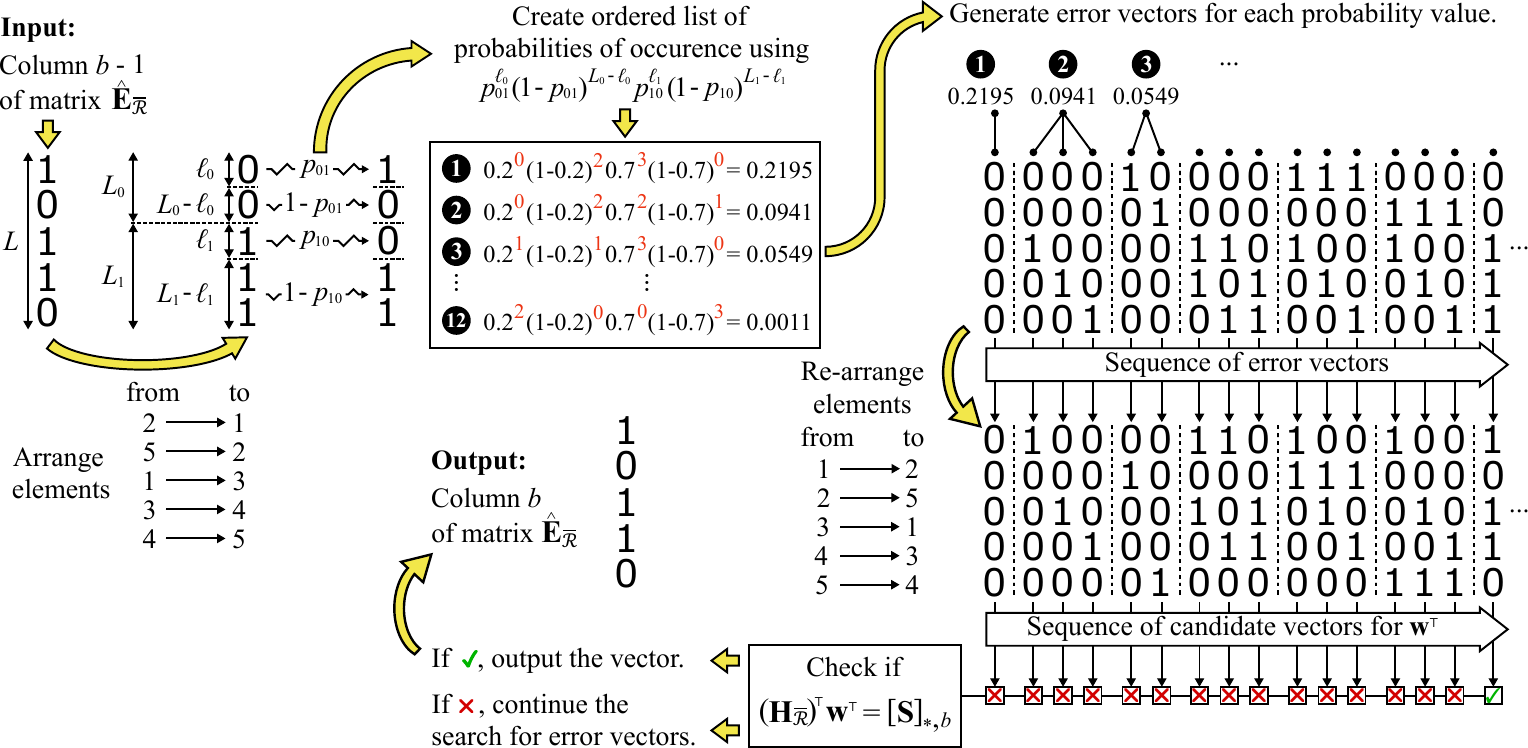}
\caption{Pictorial representation of transversal GRAND by means of an example, in which all column vectors have length $L=5$, the input vector contains $L_0=2$ zeros and $L_1=3$ ones, and the transition probabilities are $p_{01}=0.2$ and $p_{10}=0.7$.}
\label{fig:transversal_GRAND}
\end{figure*}

The transversal GRAND algorithm is illustrated in Fig.~\ref{fig:transversal_GRAND} by means of an example, in which column $b$ of $\hat{\mathbf{E}}_{\overbar{\mathcal{R}}}$, for $b=1,\ldots,B$, is estimated based on column $b-1$ of $\hat{\mathbf{E}}_{\overbar{\mathcal{R}}}$ and the transition probabilities $p_{01}$ and $p_{10}$ of the Markov model. State~$0$ is the initial state of the Markov chain, therefore the all-zero column is input to the algorithm for the estimation of the first column (i.e., $b=1$) of $\hat{\mathbf{E}}_{\overbar{\mathcal{R}}}$. The length of the input and output column vectors is $L=N-N_\mathrm{R}$.

At first, the algorithm identifies the positions of the elements of the input vector that are equal to $0$ or $1$. Let $L_0$ and $L_1$ be the number of zeros and ones, respectively. The elements of the input vector do not need to be arranged based on their values, as long as the algorithm keeps track of the positions of each value. However, we opted to arrange the elements such that the $L_0$ zeros are grouped at the top of the vector and the $L_1$ ones are grouped at the bottom of the vector, in order to facilitate the description of the algorithm. As shown in Fig.~\ref{fig:transversal_GRAND}, the $L_0=2$ zeros of the input vector are moved from positions $2$ and $5$ to positions $1$ and $2$, while the $L_1=3$ ones are moved to the three bottom positions. Essentially, the input vector contains the current states of $L$ independent Markov chains; $L_0$ of them are in state $0$ (the `good' state) and the remaining $L_1$ chains are in state $1$ (the `bad' state).

According to the Markov model, the values of $\ell_i$ of the $L_i$ elements will change from $i$ to $j$ with probability $p_{ij}$, for $i,j\in\{0,1\}$, $i\neq j$ and $\ell_i=0,\dots,L_i$. The values of the remaining $L_i-\ell_i$ elements will stay the same with probability $1-p_{ij}$. The overall probability of occurrence of the aforementioned transitions is:
\begin{equation}
p_{01}^{\ell_0}\left(1-p_{01}\right)^{L_0-\ell_0}p_{10}^{\ell_1}\left(1-p_{10}\right)^{L_1-\ell_1}.\nonumber
\end{equation}
The probability of occurrence is calculated for $\ell_0=0,\ldots,L_0$ and $\ell_1=0,\ldots,L_1$ resulting in $(L_0+1)(L_1+1)$ probability values, which are sorted in descending order. As can be seen in Fig.~\ref{fig:transversal_GRAND}, for $L_0=2$ and $L_1=3$, twelve probability values are computed and ordered. For $p_{01}=0.2$ and $p_{10}=0.7$, the input vector transitioning into the all-zero vector is the most likely event to occur, i.e., the event of the two zeros of the input vector remaining unchanged and the three ones changing to $0$ with overall probability $0.2195$.

For each value of the probability of occurrence, which has been obtained for a specific value of $\ell_0$ and $\ell_1$, the algorithm generates a sequence of $\binom{L_0}{\ell_0}\binom{L_1}{\ell_1}$ equally likely vectors. For example, the ordered list of probabilities of occurrence in Fig.~\ref{fig:transversal_GRAND} shows that the probability of $\ell_0\!=\!0$ zeros changing to $1$ and $\ell_1=2$ ones changing to $0$ is $0.0941$. The input vector, which contains $L_0=2$ zeros and $L_1=3$ ones, could thus transition to one of $\binom{2}{0}\binom{3}{2}=3$ vectors, seen in the top-right quarter of Fig.~\ref{fig:transversal_GRAND}, with probability $0.0941$. As expected, the $(L_0+1)(L_1+1)$ probability values are mapped to a total of
\begin{equation}
\sum_{\ell_0=0}^{L_0}\sum_{\ell_1=0}^{L_1}\binom{L_0}{\ell_0}\binom{L_1}{\ell_1}=\left[\sum_{\ell_0=0}^{L_0}\binom{L_0}{\ell_0}\right]\left[\sum_{\ell_1=0}^{L_1}\binom{L_1}{\ell_1}\right]=2^{L_0+L_1}=2^L\nonumber
\end{equation}
vectors. Observe in Fig.~\ref{fig:transversal_GRAND} that transversal GRAND, in contrast to syndrome decoding, does not assign by default the same probability of occurrence to vectors that have the same Hamming weight. Furthermore, the vectors that are generated by transversal GRAND are not necessarily of increasing Hamming weight. For example, the generated weight-$1$ vectors in Fig.~\ref{fig:transversal_GRAND} are followed by nine of the ten possible weight-$2$ vectors because one of the weight-$3$ vectors, that is, $[0\,0\,1\,1\,1]^\top$, is more likely to occur than the remaining weight-$2$ vector.

If the original positions of the elements of the input vector changed in the first stage of the algorithm, the elements of the generated vectors should also be re-arranged, so that they occupy their original positions. For each vector of re-arranged elements, denoted by $\mathbf{w}^\top$, the algorithm checks whether $\mathbf{w}^\top$ is a solution to the linear equation $\left(\mathbf{H}_{\overbar{\mathcal{R}}}\right)^{\!\top}\mathbf{w}^{\top}=\left[\mathbf{S}\right]_{*,b}$. As also explained in Fig.~\ref{fig:transversal_GRAND}, if $\mathbf{w}^\top$ is indeed a solution, then it is selected to be column $b$ of $\hat{\mathbf{E}}_{\overbar{\mathcal{R}}}$. Otherwise, the algorithm continues to generate and query vectors in descending order of likelihood, until a solution is found. The algorithm terminates when solutions for all $B$ linear equations, shown in \eqref{eq.syndrome_decoding_partial_per_col}, have been estimated and $\hat{\mathbf{E}}_{\overbar{\mathcal{R}}}$ has been obtained.


The following section compares syndrome decoding and transversal GRAND for different channel conditions. Note that when $p_{01}+p_{10}=1$, the Markov chain reduces to the BSC model, and transversal GRAND becomes equivalent to syndrome decoding.


\section{Results and Discussion}
\label{sec.Results}

\begin{figure*}[t!]
\centering
\subfloat[$\varepsilon\in\{0.01,0.03,0.05\}$, $\Lambda=4$ and $B=64$]{\includegraphics[width=0.666\columnwidth]{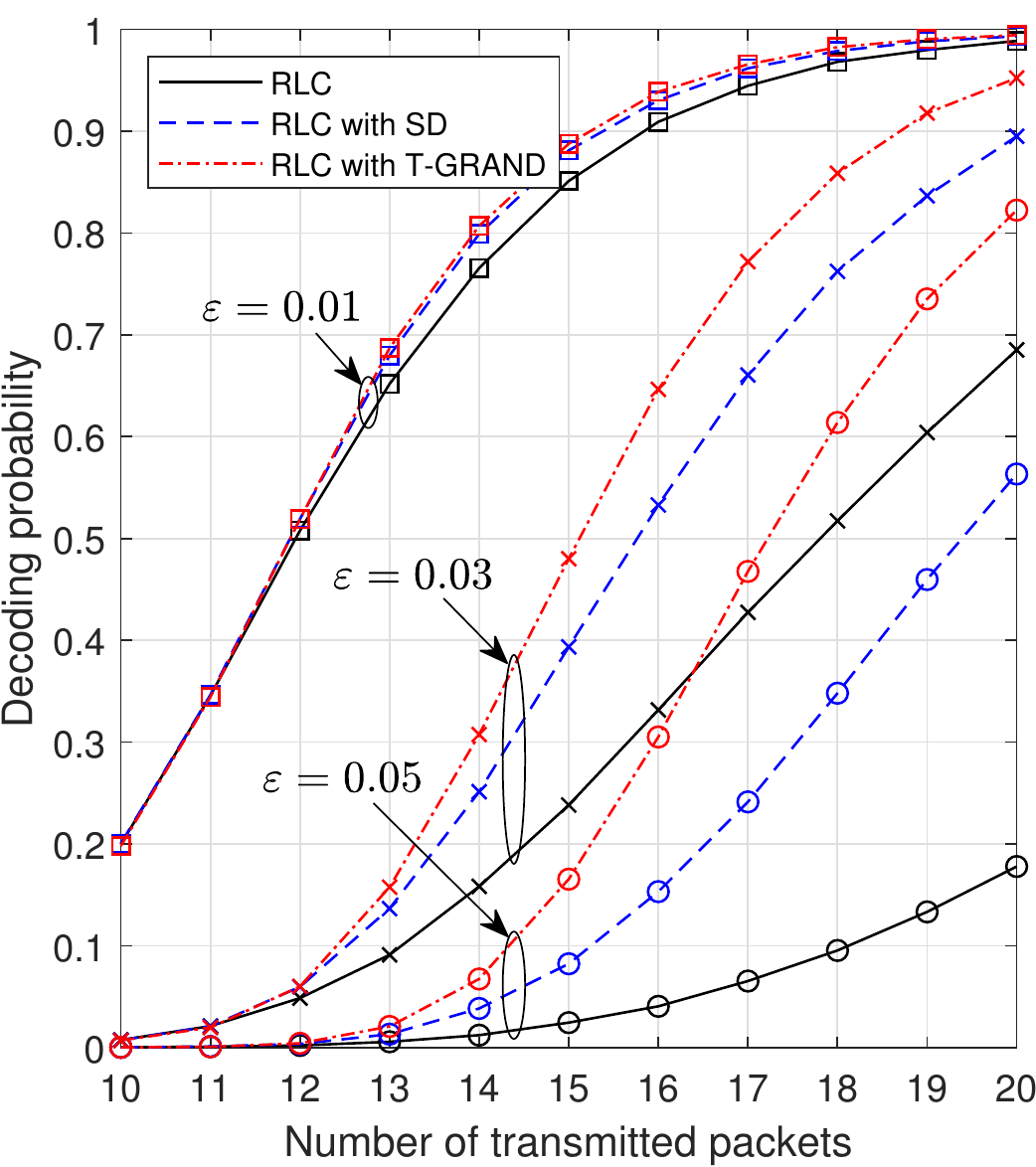}
\label{fig:var_error_prob}}
\hfil
\subfloat[$\varepsilon=0.03$, $\Lambda\in\{3,5,7\}$ and $B=64$]{\includegraphics[width=0.666\columnwidth]{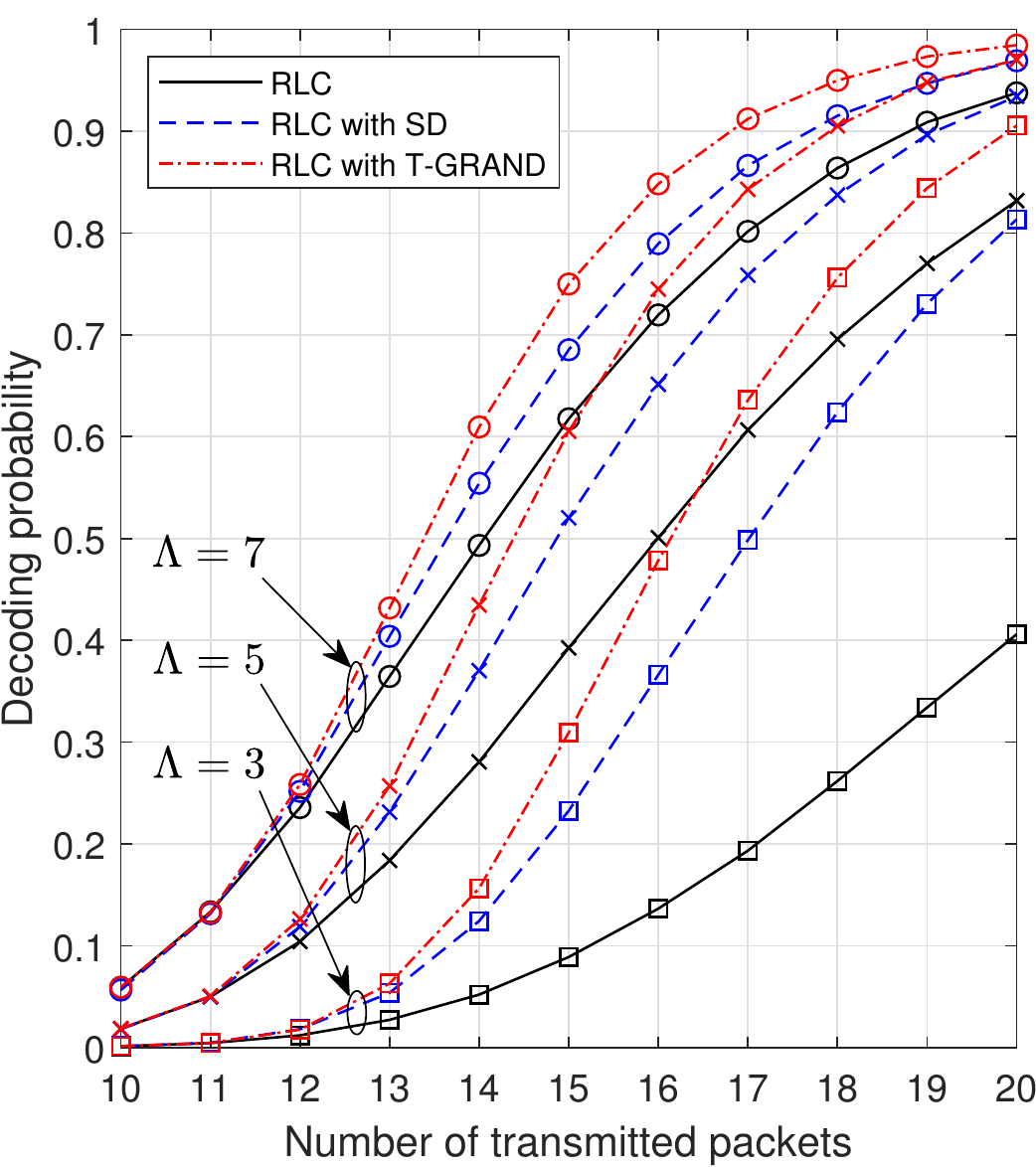}
\label{fig:var_burst_length}}
\hfil
\subfloat[$\varepsilon=0.03$, $\Lambda=3$ and $B\in\{16,32,96\}$]{\includegraphics[width=0.666\columnwidth]{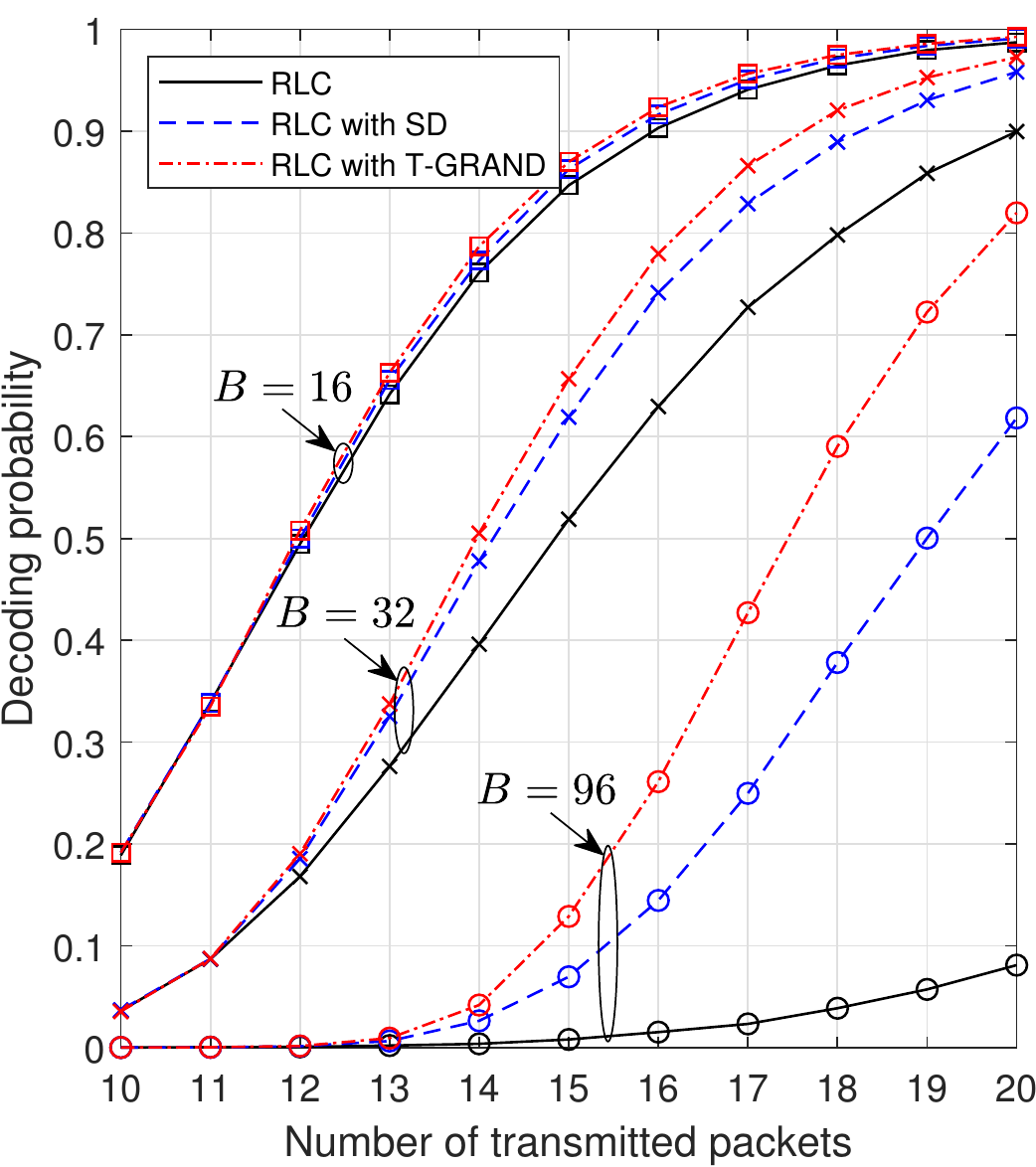}
\label{fig:var_packet_length}}
\caption{Comparison of the decoding probabilities of RLC decoding (RLC), RLC decoding combined with syndrome decoding (RLC with SD) and RLC decoding combined with transversal GRAND (RLC with T-GRAND) for $K=10$ source packets and $N$ transmitted coded packets, where $N=10,\ldots,20$. System configurations for different values of bit error probability ($\varepsilon$), average length of error bursts ($\Lambda$) and packet length ($B$) have been considered.}
\label{fig:sim_plots}
\vspace{-1pt}
\end{figure*}

In this section, the bit error probability, $\varepsilon$, and the average length of an error burst, $\Lambda$, are used as the channel parameters. For this reason, the transition probabilities, $p_{10}$ and $p_{01}$, have been expressed in terms of $\varepsilon$ and $\Lambda$ based on the definitions in Section~\ref{subsec:Gilbert}, i.e., $p_{10}=1/\Lambda$ and $p_{01}=\varepsilon/\left[\Lambda\left(1-\varepsilon\right)\right]$, and have been used as inputs to the transversal GRAND algorithm. The objective of this section is to compare the performance of stand-alone RLC decoding, RLC decoding combined with syndrome decoding (RLC with SD) and RLC decoding combined with transversal GRAND (RLC with T-GRAND) in terms of their \textit{decoding probability}, that is, the probability of a destination node recovering the $K$ source packets when $N$ coded packets have been transmitted. Measurements of the decoding probability, obtained through simulations  for $K=10$ and $N$ varying from $10$ to $20$, are presented in Fig.~\ref{fig:sim_plots}.

Fig.~\ref{fig:var_error_prob} shows measurements for an increasing probability of the channel flipping bits in transmitted packets, when each packet contains $B=64$ bits and errors occur in bursts of length $\Lambda=4$ on average. For $\varepsilon=0.01$ and a growing number of transmitted packets, the destination node is likely to receive $K$ error-free linearly independent packets among the received packets and recover the $K$ source packets using RLC decoding; SD and T-GRAND offer only a marginal improvement in decoding probability. If the value of $\varepsilon$ is raised to $0.05$, the proportion of erroneously received packets increases, the RLC decoder experiences a notable drop in decoding performance, and SD and T-GRAND become instrumental in improving the chances of recovering the source packets. For example, when $\varepsilon=0.05$ and $N=20$, SD increases the decoding probability of RLC from $0.18$ to $0.56$ whereas T-GRAND, which considers correlations in errors when repairing coded packets, boosts the decoding probability to $0.82$.

Fig.~\ref{fig:var_burst_length} depicts the impact of the average length of error bursts on the decoding probability, when $\varepsilon=0.03$. When the value of $\varepsilon$ is constant, the same number of errors -- on average -- impair the transmitted packets, but errors aggregate in fewer packets as the average length of error bursts increases. Consequently, destination nodes receive an increasingly larger proportion of error-free packets as $\Lambda$ grows, and RLC decoding stands a greater chance of success. Although fewer packets are received in error for higher values of $\Lambda$, erroneous packets are more severely damaged by longer error bursts. Nevertheless, SD and especially T-GRAND can still improve the decoding probability of RLC. For example, the decoding probability of RLC, RLC with SD and RLC with T-GRAND is $0.72$, $0.79$ and $0.85$, respectively, for $\Lambda=7$ and $N=16$. On the other hand, low values of $\Lambda$ have a negative impact on RLC decoding; for a decreasing length of error bursts and a fixed average number of errors, the proportion of correctly received packets reduces. Packets corrupted by errors may dominate but errors are more sparsely distributed, and both SD and T-GRAND are more effective in repairing corrupted packets. Notice in Fig.~\ref{fig:var_burst_length} that the decoding probability of RLC, RLC with SD and RLC with T-GRAND is $0.41$, $0.81$ and $0.91$, respectively, for $\Lambda=3$ and $N=20$.

The effect of the packet length $B$ on the decoding probability is shown in Fig.~\ref{fig:var_packet_length}. For fixed values of $\varepsilon$ and $\Lambda$, the likelihood that a packet will be corrupted by errors increases with the packet length and, thus, the performance of RLC decoding deteriorates. When it comes to SD and T-GRAND, the value of $B$ also signifies the number of times that the algorithms have run in order to estimate the errors in the $B$ positions of the corrupted packets. The advantage of T-GRAND over SD becomes apparent as the packet length increases. For $B=96$ and $N=20$, SD lifts the decoding probability of RLC from $0.08$ to $0.62$; T-GRAND further increases the decoding probability to $0.82$ as is more successful than SD in making consecutive accurate guesses of error occurrences in the course of the $B=96$ runs.


\section{Conclusions}
\label{sec.Conclusions}

This paper builds on the `guessing random additive noise decoding' (GRAND) framework and develops transversal GRAND, a hard detection decoder that leverages correlations between channel errors and works in tandem with the decoder of a random linear code (RLC). Transversal GRAND generates and queries error vectors in order of likelihood and attempts to correct bits that occupy the same position in all of the erroneously received packets. After all bit positions have been considered, RLC decoding utilizes both repaired and correctly received packets to recover the original information packets. Simulation results demonstrated that transversal GRAND has the potential to markedly improve the decoding performance of RLC when the communication channel introduces error bursts in the transmitted packets.


\IEEEtriggeratref{7}
\bibliographystyle{IEEEtran}
\bibliography{IEEEabrv,references}


\end{document}